\documentclass[twocolumn]{aastex62}

\newcommand{\Lsun}{$L_{\odot}$}
\newcommand{\Msun}{$M_{\odot}$}

\newcommand{\mic}{$\mu$m}

\newcommand{\mstar}{$M_{*}$}

\newcommand{\Lstar}{$L_{*}$}

\usepackage{xcolor}

\shorttitle{MINDS. The detection of $^{13}$CO$_{2}$ with JWST-MIRI indicates abundant CO$_{2}$ in a protoplanetary disk}
\shortauthors{Grant et al.}

\begin{document}

\title{MINDS. The detection of $^{13}$CO$_{2}$ with JWST-MIRI indicates abundant CO$_{2}$ in a protoplanetary disk}

\correspondingauthor{Sierra L. Grant}
\email{sierrag@mpe.mpg.de}

\author[0000-0002-4022-4899]{Sierra L. Grant}
\affil{Max-Planck Institut f\"{u}r Extraterrestrische Physik (MPE), Giessenbachstr. 1, 85748, Garching, Germany}

\author{Ewine F. van Dishoeck}
\affil{Leiden Observatory, Leiden University, 2300 RA Leiden, the Netherlands}
\affil{Max-Planck Institut f\"{u}r Extraterrestrische Physik (MPE), Giessenbachstr. 1, 85748, Garching, Germany}

\author{Beno\^{i}t Tabone}
\affil{Universit\'e Paris-Saclay, CNRS, Institut d’Astrophysique Spatiale, 91405, Orsay, France}

\author{Danny Gasman}
\affil{Institute of Astronomy, KU Leuven, Celestijnenlaan 200D, 3001 Leuven, Belgium}

\author[0000-0002-1493-300X]{Thomas Henning}
\affil{Max-Planck-Institut f\"{u}r Astronomie (MPIA), K\"{o}nigstuhl 17, 69117 Heidelberg, Germany}

\author{Inga Kamp}
\affil{Kapteyn Astronomical Institute, Rijksuniversiteit Groningen, Postbus 800, 9700AV Groningen, The Netherlands}

\author[0000-0001-9818-0588]{Manuel G\"udel}
\affil{Dept. of Astrophysics, University of Vienna, T\"urkenschanzstr 17, A-1180 Vienna, Austria}
\affil{Max-Planck-Institut f\"{u}r Astronomie (MPIA), K\"{o}nigstuhl 17, 69117 Heidelberg, Germany}
\affil{ETH Z\"urich, Institute for Particle Physics and Astrophysics, Wolfgang-Pauli-Str. 27, 8093 Z\"urich, Switzerland}

\author{Pierre-Olivier Lagage}
\affil{Universit\'e Paris-Saclay, Universit\'e Paris Cit\'e, CEA, CNRS, AIM, F-91191 Gif-sur-Yvette, France}

\author{Giulio Bettoni}
\affil{Max-Planck Institut f\"{u}r Extraterrestrische Physik (MPE), Giessenbachstr. 1, 85748, Garching, Germany}

\author[0000-0002-8545-6175]{Giulia Perotti}
\affil{Max-Planck-Institut f\"{u}r Astronomie (MPIA), K\"{o}nigstuhl 17, 69117 Heidelberg, Germany}

\author{Valentin Christiaens}
\affil{STAR Institute, Universit\'e de Li\`ege, All\'ee du Six Ao\^ut 19c, 4000 Li\`ege, Belgium}

\author{Matthias Samland}
\affil{Max-Planck-Institut f\"{u}r Astronomie (MPIA), K\"{o}nigstuhl 17, 69117 Heidelberg, Germany}

\author[0000-0001-8407-4020]{Aditya M. Arabhavi}
\affil{Kapteyn Astronomical Institute, Rijksuniversiteit Groningen, Postbus 800, 9700AV Groningen, The Netherlands}

\author[0000-0003-2820-1077]{Ioannis Argyriou}
\affil{Institute of Astronomy, KU Leuven, Celestijnenlaan 200D, 3001 Leuven, Belgium}

\author{Alain Abergel}
\affil{Universit\'e Paris-Saclay, CNRS, Institut d’Astrophysique Spatiale, 91405, Orsay, France}

\author[0000-0002-4006-6237]{Olivier Absil}
\affil{STAR Institute, Universit\'e de Li\`ege, All\'ee du Six Ao\^ut 19c, 4000 Li\`ege, Belgium}

\author[0000-0002-5971-9242]{David Barrado}
\affil{Centro de Astrobiolog\'ia (CAB), CSIC-INTA, ESAC Campus, Camino Bajo del Castillo s/n, 28692 Villanueva de la Ca\~nada,
Madrid, Spain}

\author{Anthony Boccaletti}
\affil{LESIA, Observatoire de Paris, Universit\'e PSL, CNRS, Sorbonne Universit\'e, Universit\'e de Paris, 5 place Jules Janssen, 92195 Meudon, France}

\author[0000-0003-4757-2500]{Jeroen Bouwman}
\affil{Max-Planck-Institut f\"{u}r Astronomie (MPIA), K\"{o}nigstuhl 17, 69117 Heidelberg, Germany}

\author{Alessio Caratti o Garatti}
\affil{INAF – Osservatorio Astronomico di Capodimonte, Salita Moiariello 16, 80131 Napoli, Italy}
\affil{Dublin Institute for Advanced Studies, 31 Fitzwilliam Place, D02, XF86 Dublin, Ireland}

\author{Vincent Geers}
\affil{UK Astronomy Technology Centre, Royal Observatory Edinburgh, Blackford Hill, Edinburgh EH9 3HJ, UK}

\author[0000-0001-9250-1547]{Adrian M. Glauser}
\affil{ETH Z\"urich, Institute for Particle Physics and Astrophysics, Wolfgang-Pauli-Str. 27, 8093 Z\"urich, Switzerland}

\author{Rodrigo Guadarrama}
\affil{Dept. of Astrophysics, University of Vienna, T\"urkenschanzstr 17, A-1180 Vienna, Austria}

\author{Hyerin Jang}
\affil{Department of Astrophysics/IMAPP, Radboud University, PO Box 9010, 6500 GL Nijmegen, The Netherlands}

\author{Jayatee Kanwar}
\affil{Kapteyn Astronomical Institute, Rijksuniversiteit Groningen, Postbus 800, 9700AV Groningen, The Netherlands}
\affil{Space Research Institute, Austrian Academy of Sciences, Schmiedlstr. 6, A-8042, Graz, Austria}

\author{Fred Lahuis}
\affil{SRON Netherlands Institute for Space Research, PO Box 800, 9700 AV, Groningen, The Netherlands}

\author{Maria Morales-Calder\'on}
\affil{Centro de Astrobiolog\'ia (CAB), CSIC-INTA, ESAC Campus, Camino Bajo del Castillo s/n, 28692 Villanueva de la Ca\~nada,
Madrid, Spain}

\author[0000-0003-3217-5385]{Michael Mueller}
\affil{Kapteyn Astronomical Institute, Rijksuniversiteit Groningen, Postbus 800, 9700AV Groningen, The Netherlands}

\author{Cyrine Nehm\'e}
\affil{Universit\'e Paris-Saclay, Universit\'e Paris Cit\'e, CEA, CNRS, AIM, F-91191 Gif-sur-Yvette, France}

\author{G\"oran Olofsson}
\affil{Department of Astronomy, Stockholm University, AlbaNova University Center, 10691 Stockholm, Sweden}

\author{Eric Pantin}
\affil{Universit\'e Paris-Saclay, Universit\'e Paris Cit\'e, CEA, CNRS, AIM, F-91191 Gif-sur-Yvette, France}

\author{Nicole Pawellek}
\affil{Dept. of Astrophysics, University of Vienna, T\"urkenschanzstr 17, A-1180 Vienna, Austria}

\author[0000-0002-2110-1068]{Tom P. Ray}
\affil{Dublin Institute for Advanced Studies, 31 Fitzwilliam Place, D02, XF86 Dublin, Ireland}

\author{Donna Rodgers-Lee}
\affil{Dublin Institute for Advanced Studies, 31 Fitzwilliam Place, D02, XF86 Dublin, Ireland}

\author[0000-0003-4559-0721]{Silvia Scheithauer}
\affil{Max-Planck-Institut f\"{u}r Astronomie (MPIA), K\"{o}nigstuhl 17, 69117 Heidelberg, Germany}

\author{J\"urgen Schreiber}
\affil{Max-Planck-Institut f\"{u}r Astronomie (MPIA), K\"{o}nigstuhl 17, 69117 Heidelberg, Germany}

\author{Kamber Schwarz}
\affil{Max-Planck-Institut f\"{u}r Astronomie (MPIA), K\"{o}nigstuhl 17, 69117 Heidelberg, Germany}

\author{Milou Temmink}
\affil{Leiden Observatory, Leiden University, 2300 RA Leiden, the Netherlands}

\author{Bart Vandenbussche}
\affil{Institute of Astronomy, KU Leuven, Celestijnenlaan 200D, 3001 Leuven, Belgium}

\author{Marissa Vlasblom}
\affil{Leiden Observatory, Leiden University, 2300 RA Leiden, the Netherlands}

\author[0000-0002-5462-9387]{L. B. F. M. Waters}
\affil{Department of Astrophysics/IMAPP, Radboud University, PO Box 9010, 6500 GL Nijmegen, The Netherlands}
\affil{SRON Netherlands Institute for Space Research, Niels Bohrweg 4, NL-2333 CA Leiden, the Netherlands}

\author{Gillian Wright}
\affil{UK Astronomy Technology Centre, Royal Observatory Edinburgh, Blackford Hill, Edinburgh EH9 3HJ, UK}

\author{Luis Colina}
\affil{Centro de Astrobiolog\'ia (CAB, CSIC-INTA), Carretera de Ajalvir, E-28850 Torrej\'on de Ardoz, Madrid, Spain}

\author{Thomas R. Greve}
\affil{DTU Space, Technical University of Denmark. Building 328, Elektrovej, 2800 Kgs. Lyngby, Denmark}

\author{Kay Justannont}
\affil{Chalmers University of Technology, Onsala Space Observatory, 439 92 Onsala, Sweden}

\author{G\"oran \"Ostlin}
\affil{Department of Astronomy, Stockholm University, AlbaNova University Center, 10691 Stockholm, Sweden}

\begin{abstract}
We present JWST-MIRI MRS spectra of the protoplanetary disk around the low-mass T Tauri star GW Lup from the MIRI mid-INfrared Disk Survey (MINDS) GTO program. Emission from $^{12}$CO$_{2}$, $^{13}$CO$_{2}$, H$_{2}$O, HCN, C$_{2}$H$_{2}$, and OH is identified with $^{13}$CO$_{2}$ being detected for the first time in a protoplanetary disk. We characterize the chemical and physical conditions in the inner few au of the GW Lup disk using these molecules as probes. The spectral resolution of JWST-MIRI MRS paired with high signal-to-noise data is essential to identify these species and determine their column densities and temperatures. The $Q$-branches of these molecules, including those of hot-bands, are particularly sensitive to temperature and column density. We find that the $^{12}$CO$_{2}$ emission in the GW Lup disk is coming from optically thick emission at a temperature of $\sim$400 K. $^{13}$CO$_{2}$ is optically thinner and based on a lower temperature of $\sim$325 K, may be tracing deeper into the disk and/or a larger emitting radius than $^{12}$CO$_{2}$. The derived $N_{\rm{CO_{2}}}$/$N_{\rm{H_{2}O}}$ ratio is orders of magnitude higher than previously derived for GW Lup and other targets based on \textit{Spitzer}-IRS data. This high column density ratio may be due to an inner cavity with a radius in between the H$_{2}$O and CO$_{2}$ snowlines and/or an overall lower disk temperature. This paper demonstrates the unique ability of JWST to probe inner disk structures and chemistry through weak, previously unseen molecular features.
\end{abstract}


\section{Introduction}\label{sec: intro}
The inner 10 au of protoplanetary disks are regions of active chemistry, with high temperatures and densities and with the snowlines of H$_{2}$O and CO$_{2}$ controlling the gas composition (e.g., \citealt{pontoppidan14a,walsh15,bosman22}). The chemistry in this region is expected to impact the atmospheric compositions of any exoplanets, the bulk of which are expected to form in this region \citep{dawson_johnson18,oberg_bergin21,molliere22}, which is difficult to probe with ALMA.

Of the several species that emit from the inner $\sim$10 au of disks, CO$_{2}$ is a particularly informative tracer of the physical and chemical conditions in this region. In the interstellar medium, ices are rich in CO$_{2}$ (abundances of 10$^{-5}$ with respect to the total gas density; \citealt{degraauw96,gibb04,bergin05, pontoppidan08,boogert15}). However, the CO$_{2}$ abundance in disks, derived from both LTE slab models and full disk non-LTE modeling of \textit{Spitzer}-IRS observations, is between 10$^{-9}$ and 10$^{-7}$ with respect to the total gas density, indicating reprocessing in the disk \citep{pontoppidan14b,salyk11a,bosman17}. Despite these lower disk abundances, \citet{pontoppidan10a} found that CO$_{2}$ was the second most common molecule detected in disks (20 disks) after water (25 disks) in a sample of 73 protoplanetary disks observed with \textit{Spitzer}-IRS. In those sources, the CO$_{2}$ $Q$-branch at 14.9 \mic\ was useful as a diagnostic of the gas temperature and abundance in their inner regions. Besides ice production, CO$_{2}$ is also formed in the gas phase at moderate temperatures (100 to 200 K) through the reaction of CO + OH $\rightarrow$ CO$_{2}$ + H. At higher temperatures, OH primarily reacts with H$_{2}$ to form H$_{2}$O. Thus, the CO$_{2}$/H$_{2}$O ratio is sensitive to the gas temperature.

Despite the usefulness of its typically bright $Q$-branch, $^{12}$CO$_{2}$ is thought to be largely optically thick in the regions of the disk where it is emitting \citep{bosman17}. Therefore, optically thinner lines and isotologues are more useful in determining the column density of CO$_{2}$, as well as the physical and chemical conditions in the disk. Moderate spectral resolution and high signal-to-noise observations are needed to detect the weaker optically thin lines and isotopologues, which was not possible with \textit{Spitzer}. JWST-MIRI provides a new opportunity to study the $Q$-branches of $^{12}$CO$_{2}$ and $^{13}$CO$_{2}$, and the ability to identify individual $P$- and $R$-branch lines for $^{12}$CO$_{2}$. 

We present JWST-MIRI observations of one of the CO$_{2}$-bright sources identified in \textit{Spitzer} observations: GW Lup \citep{pontoppidan10a,salyk11a,bosman17}. GW Lup (Sz 71) is an M1.5 star ($T_{\rm{eff}}$=3630 K, \Lstar=0.33 \Lsun, \mstar=0.46 \Msun) in the Lupus I cloud at a distance of 155 pc \citep{alcala17,andrews18}. This target was observed as part of the DSHARP survey \citep{andrews18}, which found a very narrow ring of continuum emission at a radius of 85 au in addition to a centrally-peaked continuum \citep{dullemond18}. We re-detect C$_{2}$H$_{2}$ and strong $^{12}$CO$_{2}$ emission in this disk \citep{pontoppidan10a,salyk11a,banzatti20} and additionally detect $^{13}$CO$_{2}$, H$_{2}$O, HCN, and OH for the first time in this source. We fit the 13.6 to 16.3 \mic\ wavelength range of the MIRI spectrum with local thermodynamic equilibrium (LTE) slab models to constrain the column density and temperature for each species. We discuss our findings, in particular the detection of $^{13}$CO$_{2}$, which is the first such detection in a protoplanetary disk, and the column density ratio of CO$_{2}$ to H$_{2}$O, which provides new insight into the inner disk structure.

\begin{figure*}
    \centering
    \includegraphics[scale=0.57]{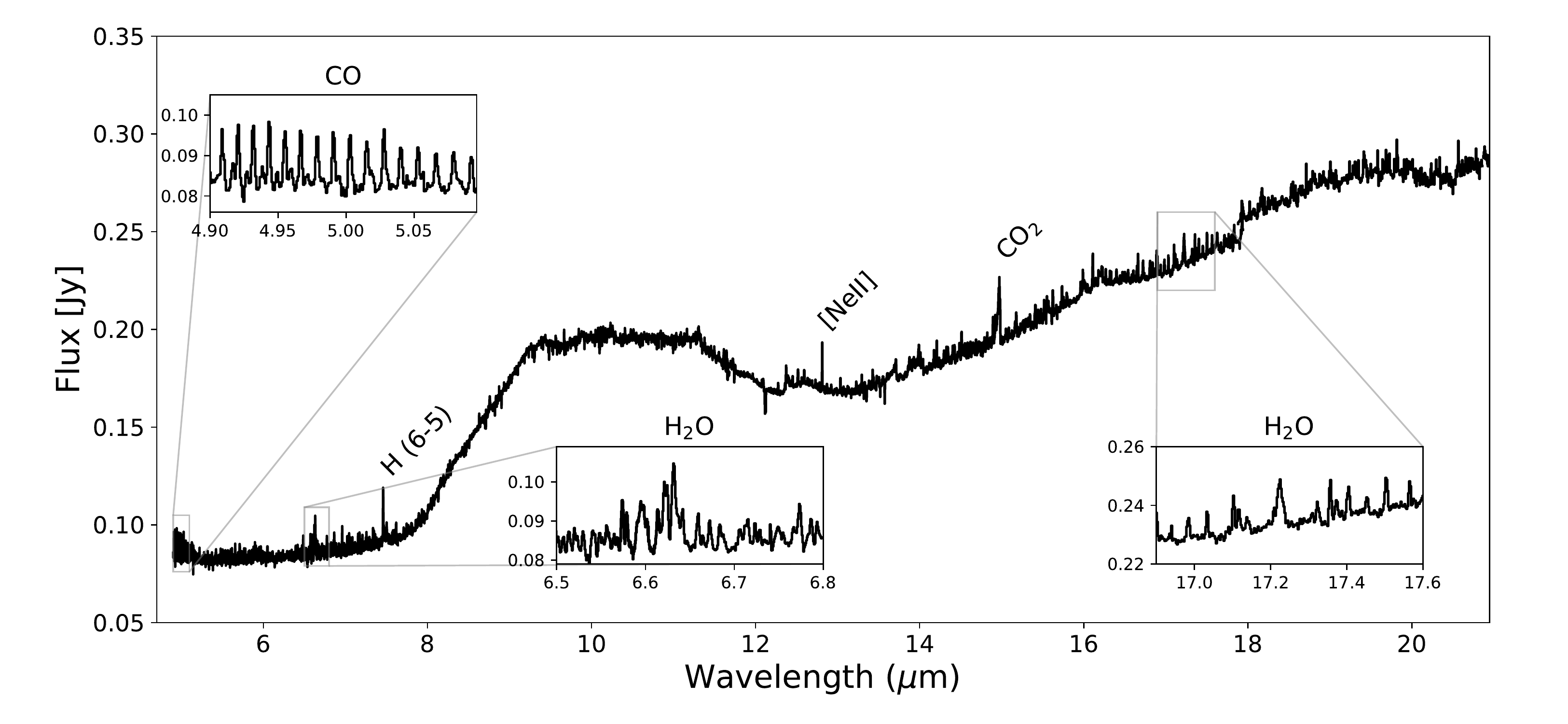}
    \caption{The JWST-MIRI MRS spectrum for GW Lup through Channel 4A. Several of the strongest emission features are labeled and insets show additional molecular features. The beginning and end of each sub-band has been trimmed to reduce spurious features due to the increased noise at the ends of the bands. }
    \label{fig: spec}
     
\end{figure*}

\section{Observations and analysis}\label{sec: observations}

\subsection{Observations and data reduction}\label{subsec: obs}
GW Lup was observed with the Mid-InfraRed Instrument (MIRI; \citealt{rieke15, wells15, wright15}, Wright et al. submitted, Argyriou et al. in prep.) in the Medium Resolution Spectroscopy (MRS) mode on 8 August, 2022. These observations are part of the MIRI mid-INfrared Disk Survey (MINDS) JWST GTO Program (PID: 1282, PI: T. Henning). Target acquisition was used so that a point-source fringe flat could be used in the data reduction. A four-point dither was performed in the positive direction. The total exposure time was 1 hour. All the JWST data used in this paper can be found in MAST: \dataset[10.17909/aez2-za93]{http://dx.doi.org/10.17909/aez2-za93}.

The MIRI MRS observations were processed through all three reduction stages \citep{bushouse22} using Pipeline version 1.8.4. The reference files were generated from the observation of the reference A-type star HD 163466. Additional details on the reference files used can be found in Gasman et al. (in press). A single dedicated point-source fringe flat and dedicated spectrophotometric calibration were used in the reduction process. We skip the outlier rejection step in \texttt{Spec3}, as this produces spurious results due to the under-sampling of the PSF, causing under-sampling artefacts (e.g., short-period oscillations at the beginning and end of each sub-band) in the extracted spectrum. Under-sampling of the PSF and its artefacts will be discussed in an upcoming paper (Law et al. in preparation). Finally, the centroid of the PSF was found manually prior to the extraction of the spectra in each sub-band which included aperture correction, with an aperture size of 2.5$\lambda$/$D$. The correction factors are the same as those presented in Argyriou et al. (in preparation), which include the contribution of the PSF wings to the estimated background determined from an annulus around the source. The background emission from this annulus is subtracted and its value ranges from $\sim$0.001 Jy in Channel 1 to $\sim$0.1 Jy around 22 \mic. The continuum emission is not extended in the 2D images, therefore the disk is not resolved.

The final GW Lup spectrum through Channel 4A is presented in Figure~\ref{fig: spec}. At longer wavelengths the flux calibration becomes increasingly uncertain, due to the low flux level at these wavelengths in the reference star used for calibration, therefore we only show through Channel 4A. The $\nu_{2}$=1-0 $^{12}$CO$_{2}$ $Q$-branch is the most prominent, but a large number of weaker lines are detected as well. A spurious, single-pixel spike at 18.8 \mic\ has been removed. Below 7.5 \mic, the \textit{Spitzer} low-resolution spectrum of GW Lup has a $\sim$20\% higher flux. However, above 7.5 \mic, the flux of the JWST-MIRI spectrum is $\sim$15\% higher than the \textit{Spitzer}-IRS high- and low- resolution spectra of this target, but the overall shape is very similar (see Figure~\ref{fig: subbands and spitzer} in the Appendix). These offsets may be due to calibration issues and/or variability in this system. This will be investigated in a future work. In this work, we use the MIRI continuum-subtracted spectrum for analysis.

\begin{figure*}
    \centering
    \includegraphics[scale=0.5]{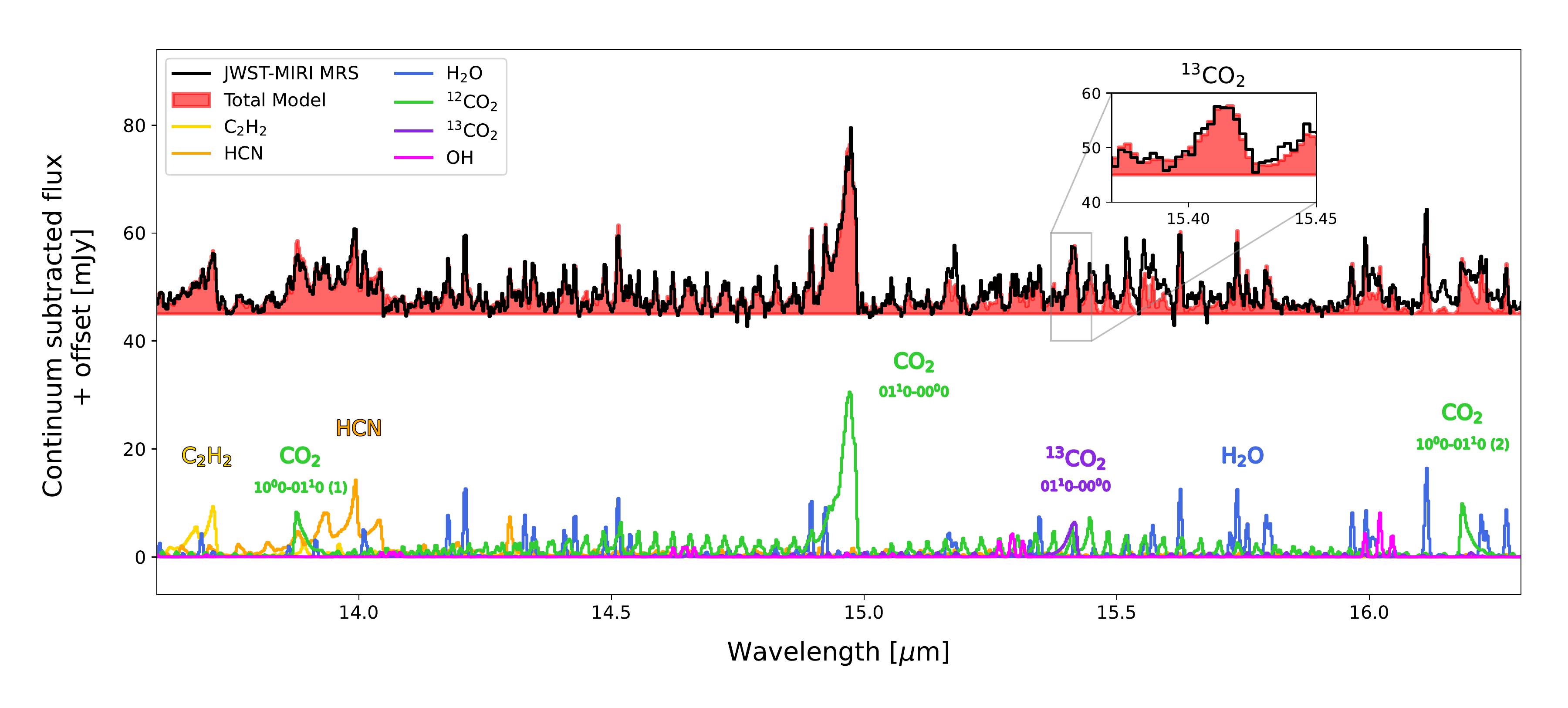}
    \caption{The 13 to 16.3 \mic\ wavelength range of the GW Lup spectrum, with the JWST-MIRI data (black) compared to a model (red) composed of emission from C$_{2}$H$_{2}$ (yellow), HCN (orange), H$_{2}$O (blue), $^{12}$CO$_{2}$ (green), $^{13}$CO$_{2}$ (purple), and OH (pink). The inset shows a zoom-in of the $^{13}$CO$_{2}$ feature. 
    }
    \label{fig: spec and model}
    
\end{figure*}

\begin{figure*}
    \centering
    \includegraphics[scale=0.58]{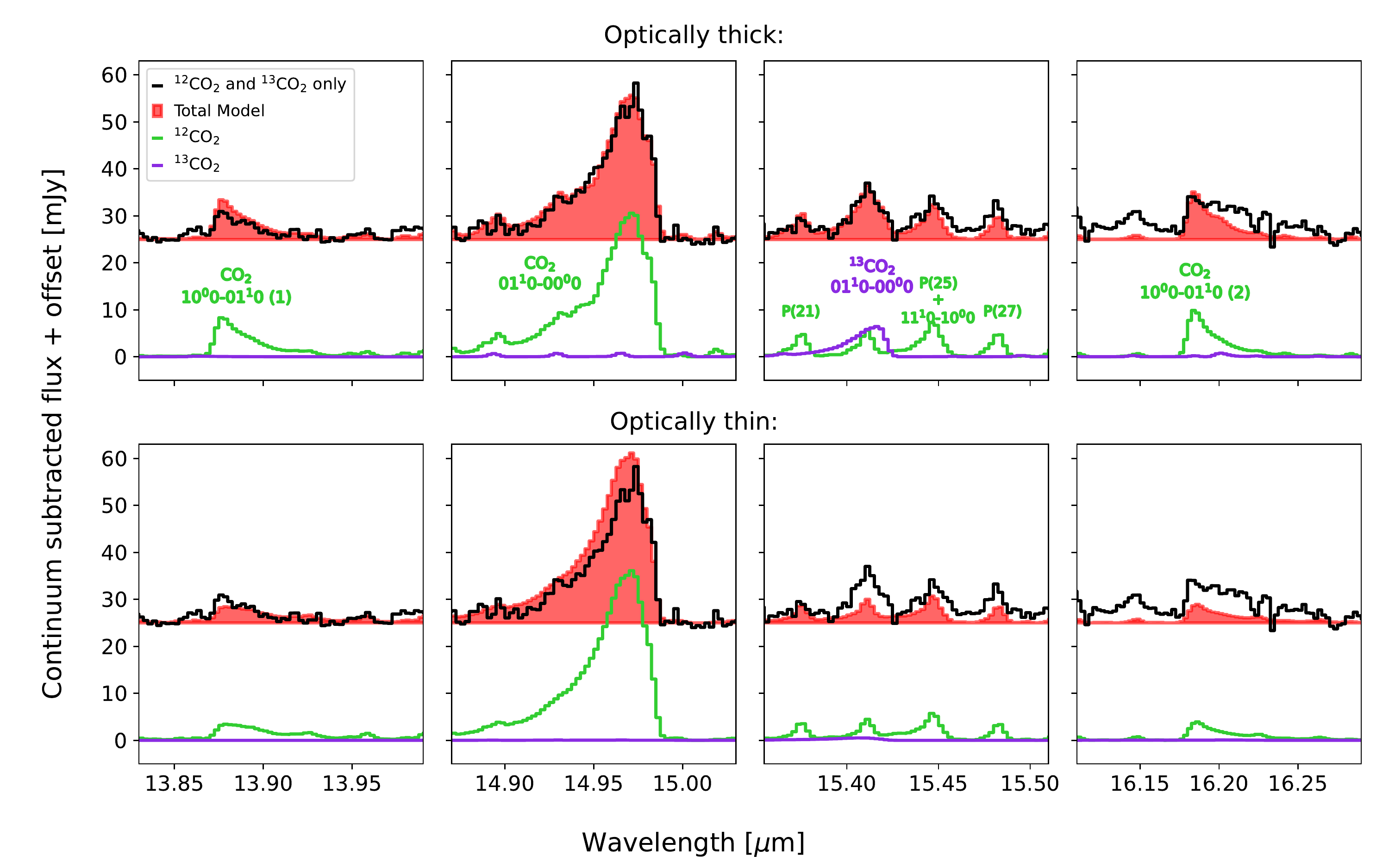}
    \caption{Zoom-ins of the 15 \mic\ wavelength range of GW Lup, with the JWST-MIRI data (black) compared to a model (red) composed of emission from $^{12}$CO$_{2}$ (green) and $^{13}$CO$_{2}$ (purple). The continuum and the best-fit models of C$_{2}$H$_{2}$, HCN, H$_{2}$O, and OH shown in Figure~\ref{fig: spec and model}, have been subtracted from the GW Lup spectrum. The top row shows the best optically thick fit from Figure~\ref{fig: spec and model}, while the bottom row shows the optically thin $^{12}$CO$_{2}$ fit from \cite{salyk11a}. The optically thick model reproduces the $^{12}$CO$_{2}$ hot-band $Q$-branches at 13.9 and 16.2 \mic\ better than the optically thin model. In the optically thin case, a $^{12}$CO$_{2}$/$^{13}$CO$_{2}$ ratio of 68 is not able to reproduce the $^{13}$CO$_{2}$ $Q$-branch at 15.4 \mic. 
    }
    \label{fig: 4 panel}
    
\end{figure*}

\subsection{Slab Modeling Procedure}\label{subsec: slab models}
The MIRI spectrum is continuum subtracted in the 13 to 17 \mic\ range by selecting regions with minimal line emission and using a cubic spline interpolation to determine the continuum level. This continuum is then subtracted from the observed spectra (see Section~\ref{sec: continuum subtraction} in the Appendix and Figure~\ref{fig: continuum subtraction} for more details). 

We fit the 13 to 16.3 \mic\ continuum-subtracted spectrum with local thermodynamic equilibrium (LTE) slab models. The line profile function is assumed to be Gaussian with a full width at half maximum of $\Delta$V=4.7 km s$^{-1}$ ($\sigma$=2 km s$^{-1}$) as is done in \cite{salyk11a}, which represents the line width for H$_2$ at 700 K. This value does not have a large impact on the results and we adopt the value of \cite{salyk11a} for consistency. The model takes into account the mutual shielding of adjacent lines for the same species. In particular, the total opacity is first computed on a fine wavelength grid by summing the contribution of all the lines before computing the emerging line intensity. These models allow us to reproduce the data with only three free parameters: the line-of-sight column density $N$, the gas temperature $T$, and the emitting area given by $\pi$$R^{2}$ for a disk of emission with radius $R$. While we report the emitting area in terms of this emitting radius, the emission could be coming from a ring with an area equivalent to $\pi$$R^{2}$. We include emission from C$_{2}$H$_{2}$, HCN, H$_{2}$O, $^{12}$CO$_{2}$, $^{13}$CO$_{2}$, and OH. The $^{12}$CO$_{2}$, $^{13}$CO$_{2}$, C$_{2}$H$_{2}$, and HCN line transitions are derived from the HITRAN database \citep{gordon22}. All the lines within 4-30 \mic\ range are selected and are converted into LAMDA format \citep{vanderTak20} for compatibility with our slab model. The partition sums for these molecules are retrieved from the \texttt{TIPS$\_$2021$\_$PYTHON} package provided by the database\footnote{https://hitran.org/suppl/TIPS/TIPS2021/}. The OH spectroscopy stems from \cite{tabone21} who used data from \cite{yousefi18} and \cite{brooke16}. We vary the emitting area, as described below, and compute a synthetic spectrum in Jy, assuming a distance to GW Lup of 155 pc. The model spectrum is then convolved to a resolving power of 2500 for Channel 3 where the emission features are present \citep{labiano21}. Finally, the convolved model spectrum is resampled to have the same wavelength grid as the observed spectrum.

For each molecule, a grid of models was run with $N$ from 10$^{14}$ to 10$^{22}$ cm$^{-2}$, in steps of 0.166 in log$_{10}$-space, and $T$ from 100 to 1500 K, in steps of 25 K. The emitting area is varied by ranging the radius from 0.01 to 10 au in steps of 0.03 in log$_{10}$-space. The best-fit $N$ and $T$ are determined using a $\chi^{2}$ fit (see Appendix~\ref{sec: chi2 procedure and maps} and Figure~\ref{fig: chi2 maps} for more details) between the continuum subtracted data and the convolved and resampled model spectrum. For each $N$ and $T$, the best-fit emitting area is determined by minimizing the $\chi^2$. The $\chi^{2}$ fit is done in spectral windows which are selected to minimize the contribution of emission from other species while still containing features that help to constrain the fits (e.g., optically thin lines and line peaks that are sensitive to temperature). This is done in an iterative approach to further reduce contamination from other species. The best-fit model is found for H$_{2}$O first. This model is then subtracted from the observed, continuum-subtracted spectrum. We then fit HCN, subtract that model, and continue that procedure for C$_{2}$H$_{2}$, $^{12}$CO$_{2}$, $^{13}$CO$_{2}$, and then the final fit is done for OH. This is illustrated, with the spectral windows used for the $\chi^2$ determinations, in Figure~\ref{fig: fit procedure} in Appendix~\ref{sec: chi2 procedure and maps}. After the initial best-fit models are found, this process is repeated for each molecule, after subtracting the best-fit models for all other species. For instance, the best-fit models for HCN, C$_{2}$H$_{2}$, $^{12}$CO$_{2}$, $^{13}$CO$_{2}$, and OH are subtracted from the observed spectra before the best-fit H$_2$O model is found again (Figure~\ref{fig: fit procedure 3}). We repeat this process a third time, after which we see no further improvements in the residuals. The $\chi^{2}$ maps are shown in Appendix \ref{sec: chi2 procedure and maps} (Figure~\ref{fig: chi2 maps}). 

\section{Results}\label{sec: results}
The best-fit model is presented in Figure~\ref{fig: spec and model} together with the continuum-subtracted spectrum in the 13.6 to 16.3 \mic\ range. The best-fit model parameters are given in Table~\ref{tab: model values}. The $\chi^{2}$ maps show that at low column densities (below $\sim$10$^{17}$ to 10$^{18}$ cm$^{-2}$, depending on the molecule, see Figure~\ref{fig: chi2 maps}), in the optically thin regime, the column density and emitting radius are completely degenerate (e.g., \citealt{salyk11a}). In the optically thick regime, the emitting radius can more accurately be determined, although there is still a degeneracy between temperature and column density. Typical uncertainties can be read from the $\chi^2$ maps where the degeneracies are also evident. The degeneracy between our three free parameters is reduced by fitting a combination of optically thick and thin lines. For optically thin emission, the total number of molecules $\mathcal{N}_{tot}$ = $\pi$$N$$R^2$ is well determined and is included in Table~\ref{tab: model values}. In GW Lup, we find that the emission of all species is optically thick, or at least on the border between optically thick and optically thin, therefore the number of molecules should be taken as a lower limit.

\subsection{$^{12}$CO$_{2}$ and $^{13}$CO$_{2}$}
Our best-fitting $^{12}$CO$_{2}$ model has $N$=2.2$\times$10$^{18}$ cm$^{-2}$, a temperature of 400 K, and an emitting radius of 0.11 au. $^{12}$CO$_{2}$ is well constrained to a temperature below $\sim$700 K, with the shape of the main $Q$-branch being particularly constraining for the temperature. Using similar models and a similar technique on \textit{Spitzer}-IRS data, \cite{salyk11a} fit the fundamental $^{12}$CO$_{2}$ $Q$-branch at 14.9 \mic\ and find a temperature of 750 K, a column density of 1.6$\times$10$^{15}$ cm$^{-2}$, and an emitting radius of 1.01 au. While this optically thin model from \cite{salyk11a} reproduces the main 14.9 \mic\ $Q$-branch, it does not reproduce the $^{12}$CO$_{2}$ hot-band $Q$-branches at 13.9 \mic\ and 16.2 \mic\ (10$^{0}$0-01$^{1}$0) as well as the optically thick model does (Figure~\ref{fig: 4 panel}). An example of the effect of changing the temperature by $\pm$ 100 K and column density by $\pm$ 0.5 dex on the $^{12}$CO$_2$ model is shown in the Appendix in Figure~\ref{fig: model example}. 

For $^{13}$CO$_{2}$, models from both the optically thick and optically thin regimes reproduce the feature well. With respect to $^{12}$CO$_{2}$, the standard $^{12}$CO$_{2}$/$^{13}$CO$_{2}$ abundance ratio of 68 from the local interstellar medium \citep{wilson_rood94, milam05} is within the allowable range. The $^{13}$CO$_{2}$ temperature is lower than that of $^{12}$CO$_{2}$, with the best-fit temperature of 325 K. This indicates that the optically thinner $^{13}$CO$_{2}$ is tracing deeper layers into the disk or larger radii (e.g., in a thin annulus farther out than the $^{12}$CO$_2$, but with the same emitting area). The combination of $^{13}$CO$_{2}$ and the P(23) line of $^{12}$CO$_{2}$, reproduces the feature at 15.42 \mic.

\subsection{H$_{2}$O}
We include emission from both para- and ortho- water, assuming ortho/para = 3 (e.g., \citealt{vandishoeck21} and references therein). Many H$_{2}$O lines are present in the 13 to 17 \mic\ region in the GW Lup spectra, however they are weaker than the main CO$_{2}$ $Q$-branch and were not seen previously by \textit{Spitzer}. An LTE slab model with a temperature of 625 K, a column density of 3.2$\times$10$^{18}$ cm$^{-2}$, and an emitting radius of 0.15 au reproduces the lines in this region. This similar H$_{2}$O column density compared to CO$_{2}$ is in contrast to the much lower CO$_{2}$/H$_{2}$O ratios found in the large T Tauri \textit{Spitzer} sample by \cite{salyk11a}, which is discussed in Section~\ref{sec: discussion}.

\subsection{Other species}
For C$_{2}$H$_{2}$ and HCN (including the 02$^{0}$0-01$^{1}$0 HCN hot-band $Q$-branch at 14.3 \mic), the fits point to temperatures of $\sim$500 K and $\sim$875 K, respectively, however this is quite unconstrained for C$_2$H$_2$, in particular. This high HCN temperature is needed to reproduce the ratio of line peaks in the main $Q$-branch. The column densities for C$_{2}$H$_{2}$ and HCN are both on the border between being optically thick and optically thin. The emitting area for C$_{2}$H$_{2}$ and HCN is chosen from the best-fit model, but it is not well constrained. The OH emission is weak in the GW Lup spectrum, leading to quite unconstrained parameters, however it is clear that the temperature is high ($\gtrsim$1000 K). OH levels are likely out of thermal equilibrium with an excitation temperature set by non-thermal processes such as prompt emission \citep{carr_najita14,tabone21} or chemical pumping \citep{liu00}.

\begin{table}
\centering
\caption{Best-fit model parameters}
\label{tab: model values}
\begin{tabular}{ccccccc}
\hline \hline 
Species & $N$ & $T$ & $R$ & $\mathcal{N}_{tot}$\footnote{As the best-fit model parameters reported here are either totally in the optically thick regime or on the border between optically thick and thin, the total molecule number should be taken as a lower limit.}\\
 &   [cm$^{-2}$] &  [K] & [au] & [mol.] \\
\hline
H$_{2}$O & 3.2$\times$10$^{18}$ & 625 & 0.15 & 5$\times$10$^{43}$ \\
HCN & 4.6$\times$10$^{17}$ & 875 & 0.06 & 1.2$\times$10$^{42}$ \\
C$_{2}$H$_{2}$ & 4.6$\times$10$^{17}$ & 500 & 0.05 & 9.3$\times$10$^{41}$ \\
$^{12}$CO$_{2}$ & 2.2$\times$10$^{18}$ & 400 & 0.11 & 1.7$\times$10$^{43}$\\
$^{13}$CO$_{2}$ & 1$\times$10$^{17}$ & 325 & 0.11 & 9.3$\times$10$^{41}$\\
OH & 1$\times$10$^{18}$ & 1075 & 0.06 & 2.6$\times$10$^{42}$\\
\hline 
\end{tabular}
\end{table}

\section{Discussion}\label{sec: discussion}

As CO$_{2}$ and H$_{2}$O are two of the main oxygen carriers in protoplanetary disks, the relative abundances of these species is informative. While H$_{2}$O emission is present in the MIRI MRS spectrum of GW Lup, it is relatively weak compared to CO$_{2}$ with an $N_{\rm{CO}_{2}}$/$N_{\rm{H_{2}O}}$ ratio of $\sim$0.7. 
It should be stressed that column density ratios should not be equated with abundance ratios since the emission of different molecules (or even of different bands of the same molecule) may originate from different regions or layers of the disk \citep{bruderer15,woitke18b}. Moreover, the emission seen at mid-infrared wavelengths only probes the upper layers of the disk above the $\tau_{\rm mid-IR}=1$ contour where the dust continuum
becomes optically thick. To infer local abundances and their ratios, retrieval methods such as used in \cite{mandell12} or full forward thermochemical models using a physical structure tailored to the GW Lup disk
are needed. Such models are beyond the scope of this paper. However, given the relatively small difference in emitting radii between CO$_2$ and H$_2$O, it is still informative to put the column density ratio into perspective.

The large T Tauri \textit{Spitzer} sample of \cite{salyk11a} find a median $N_{\rm{CO}_{2}}$/$N_{\rm{H_{2}O}}$ ratio of 5$\times$10$^{-4}$. However, the column density ratios from \cite{salyk11a} are largely derived from the low column density and high temperature (optically thin) regime that we exclude using the weaker hot-band $^{12}$CO$_{2}$ Q-branches at 13.9 and 16.2 \mic. If the best-fitting radii found for H$_2$O and CO$_2$ of 0.15 and 0.11 au, respectively, correspond to the actual emitting radius (i.e., not coming from a thin annulus at larger radii with the same emitting area), then we note that these radii are smaller than the estimated midplane snowline of H$_{2}$O, which is at $\sim$0.3-0.4 au for the stellar mass of GW Lup \citep{mulders15}. At high temperatures greater than $\sim$250 K, OH will react with H$_{2}$ to form H$_{2}$O \citep{glassgold09}. Why then is CO$_{2}$ so abundant relative to H$_{2}$O at these temperatures and locations in this disk? We present three scenarios.

\begin{itemize}

    \item Temperature structure: The temperature structure of the disk, which is largely controlled by the stellar luminosity, has a large impact on the inner disk CO$_2$ and H$_2$O abundances and on their molecular emission (e.g., \citealt{walsh15,woitke18b, anderson21}). Models and observations have shown that the inner disks around low-mass stars are richer in carbon-bearing species in the disk atmosphere than those around higher-mass stars (e.g., \citealt{pascucci13, walsh15}). GW Lup, with a stellar mass of 0.46 \Msun, may be a borderline case where the C/O in the infrared emitting region is moderately high, but not so high that C$_2$H$_2$ is booming. Additionally, it is clear that H$_2$O is not so abundant in the upper layers that self-shielding is taking place, since $^{13}$CO$_{2}$ is detected, indicating a deep layer of CO$_{2}$. \cite{bosman22} show that this occurs if the vertical H$_2$O column density remains low enough that water self-shielding is suppressed, producing more OH which is needed for additional CO$_{2}$ formation. There is always some small amount of H$_2$O in the disk atmosphere that is being dissociated by UV photons from the star, decreasing the water abundance. This may be contributing to the relatively weak H$_2$O emission as some OH emission is detected. In the cooler disks around M-type stars, some of the oxygen may also be driven into the unobservable O$_2$ rather than H$_2$O, making the atmosphere appear to be carbon rich even though the C/O ratio is solar \citep{walsh15}. The moderately low luminosity of GW Lup may contribute to the the higher $N_{\rm{CO}_{2}}$/$N_{\rm{H_{2}O}}$ determined here, however GW Lup is not so low-mass that this is likely the sole explanation. Future observations of additional targets, will help to demonstrate the impact of temperature structure on the derived column densities.

    \item Pebble drift: If dust grains coated in CO$_2$-rich ices are drifting inward from the outer disk without any traps halting the drift, the inner disk will be enriched in oxygen \citep{banzatti20}. However, if ice enrichment is taking place both CO$_{2}$ and H$_{2}$O should be enriched at a ratio of 0.2 to 0.3 \citep{boogert15}, but only if the ices are transported vertically and there is no chemical reset. There may still be additional H$_{2}$O and CO$_{2}$ hidden below the dust $\tau_{15 \mu m}$=1 line. \cite{bosman17} compare the flux of the $^{13}$CO$_{2}$ $Q$-branch at 15.42 \mic\ to the neighboring $^{12}$CO$_{2}$ P(25) line at 15.45 \mic, and show that this ratio is sensitive to enhancements of CO$_{2}$ at its snowline. In GW Lup, the P(25) line is stronger than the P(23) line, indicative of a contribution from the 11$^{1}$0-10$^{0}$0 hot-band $Q$-branch of $^{12}$CO$_{2}$ that is only present at high temperatures/column densities and was not seen in the models of \cite{bosman17}. Therefore, the P(25) line is not a good representative of an individual $P$-branch line at these $J$-levels. Instead, the P(27) line at 15.48 \mic\ can be used. From the best-fit models, the peak of the $^{13}$CO$_{2}$ $Q$-branch is at 6.5 mJy, compared to 4.7 mJy for the P(27) line. In the modeling setup of \cite{bosman17}, this $^{13}$CO$_{2}$/P(27) line ratio of 1.4 points to a low outer (10$^{-8}$ with respect to the total gas density) and high inner (10$^{-6}$) disk CO$_{2}$ abundance. While this is intriguing, further modeling efforts, such as the full 2D physical-chemical models used by \cite{bosman17} are needed to realistically compare the inner and outer disk CO$_2$ distribution for GW Lup specifically.

    \item Inner cavity and/or dust trap: If there is an inner gas and dust cavity in the disk that extends to between the H$_{2}$O and CO$_{2}$ snowlines (estimated at $\sim$0.4 and $>$1 au, respectively), the H$_{2}$O will be suppressed but the CO$_{2}$ will still be abundant in the gas phase. The models of T Tauri disks from \cite{walsh15} and \cite{anderson21} show that the column densities of H$_2$O dominate over those of CO$_2$ in the inner 1 au. A cavity or gap may be present in GW Lup, which would remove abundant H$_2$O and result in the relatively strong CO$_2$ lines and relatively weak H$_2$O lines observed in the GW Lup disk. In this case, the H$_{2}$O will only be present in the uppermost layers of the disk atmosphere or at the heated edge of the cavity/gap where the icy grains are warm enough to sublimate water ice and/or where any free volatile oxygen is driven into H$_{2}$O. \cite{anderson21} find that the H$_2$O flux decreases substantially if an inner gas cavity is present, while the CO$_2$ flux is less affected due to having more contribution from emission at larger radii, although as they note, the fluxes may not accurately reflect the column densities and abundances. If there is a dust trap between the snowlines, either created by the same mechanism opening the cavity or by other means, water ice-rich grains could be trapped beyond the H$_2$O snowline, keeping H$_2$O from sublimating but allowing for the sublimation and enrichment of CO$_{2}$. Such a small cavity cannot be seen in the ALMA data, even with the high resolution of the DSHARP data. A small cavity in the dust could be traced in the dust continuum with near-infrared interferometry, whereas a gas cavity could be traced using high-spectral resolution spectroscopy, for instance, of the CO ro-vibrational lines at 4.7 \mic\ (e.g., \citealt{brown13,banzatti22a}). None of these data exist yet for the GW Lup disk.
    
\end{itemize}

\begin{figure}
    \includegraphics[scale=0.55]{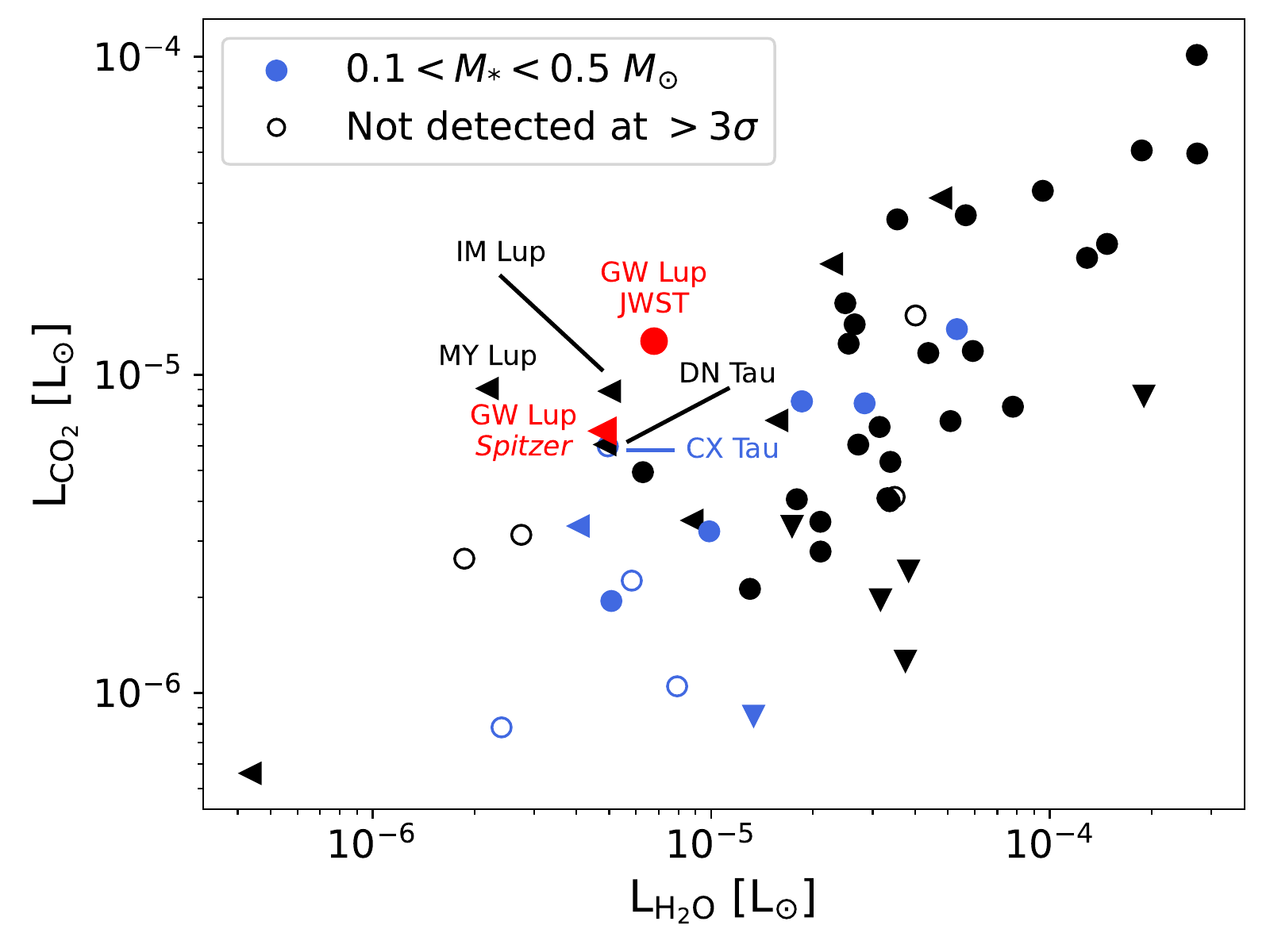}
    \caption{ The \textit{Spitzer} CO$_{2}$ $Q$-branch vs. 17 \mic\ H$_{2}$O luminosities in the sample studied in \cite{banzatti20}. Downward-facing triangles are those with CO$_{2}$ not detected above 3$\sigma$, left-ward facing triangles are those with H$_{2}$O not detected above 3$\sigma$, and open points are those with neither detected above 3$\sigma$. Blue points are stars with stellar masses between 0.1 and 0.5 \Msun, to be comparable to GW Lup (red) which has a stellar mass of 0.46 \Msun. We show the luminosities determined from \textit{Spitzer} and JWST for GW Lup.  
    }
    \label{fig: banzatti}
\end{figure}

As more sources are observed with JWST-MIRI, these scenarios can be explored further. For instance, looking for trends of the CO$_2$ vs. H$_2$O as a function of stellar luminosity and outer disk dust radius will be very informative in distinguishing between the importance of temperature structure vs. pebble drift, as was done with \textit{Spitzer} data. In the meantime, GW Lup can be put into context with other sources based on the \textit{Spitzer} fluxes. \cite{banzatti20} (re-)determined molecular line fluxes for H$_{2}$O, HCN, C$_{2}$H$_{2}$, and CO$_{2}$ for the \textit{Spitzer} sample. While GW Lup has a relatively high $Q$-branch CO$_{2}$ flux and relatively low 17 \mic\ H$_{2}$O flux compared to other disks, it is not a complete outlier (converted to line luminosities for comparison; Figure~\ref{fig: banzatti}). Several disks in the \textit{Spitzer} survey analyzed by \cite{pontoppidan10a}, \cite{salyk11a}, and \cite{banzatti20} also show high CO$_{2}$ fluxes and low water fluxes, including DN Tau, IM Lup, MY Lup, and CX Tau (Figure~\ref{fig: banzatti}). With the sensitivity and resolution of MIRI, there are likely to be other disks that will show $^{13}$CO$_{2}$ emission, which will allow us to derive strong constraints on the CO$_{2}$ column density and $N_{\rm{CO}_{2}}$/$N_{\rm{H_{2}O}}$ ratio in these disks.

\section{Summary and Conclusions}

\begin{enumerate}
    \item We identify $^{12}$CO$_{2}$, $^{13}$CO$_{2}$, H$_{2}$O, HCN, C$_{2}$H$_{2}$, and OH in the JWST-MIRI spectrum of GW Lup. Using LTE slab models, we reproduce the 13.6 to 16.3 \mic\ spectrum. H$_{2}$O, HCN, $^{13}$CO$_{2}$, and OH are detected for the first time in this disk, as the features had line/continuum ratios that were too low to be detectable at the spectral resolution of \textit{Spitzer}-IRS. 
    
    \item The gas-phase $^{13}$CO$_{2}$ detection is the first in a protoplanetary disk. This detection points to a high CO$_{2}$ abundance deep into the disk, with a $^{12}$CO$_{2}$ column density of 2.2$\times$10$^{18}$ cm$^{-2}$, temperature of 400 K, and an emitting radius of 0.11 au. For $^{13}$CO$_{2}$, the best-fit model has a column density of 1$\times$10$^{17}$ cm$^{-2}$, temperature of 325 K, and an emitting radius of 0.11 au.

    \item The column density ratio of CO$_{2}$ to H$_{2}$O derived from LTE slab models ($N_{\rm{CO}_{2}}$/$N_{\rm{H_{2}O}}\sim$0.7) that fit simultaneously the $^{12}$CO$_{2}$ hot-bands, is over two orders of magnitude higher than what has previously been found in typical T Tauri disks. This may indicate an inner cavity with a radius in between the H$_{2}$O and CO$_{2}$ midplane snowlines and/or an overall lower disk temperature. 
    \item While GW Lup has a high CO$_{2}$ flux relative to H$_{2}$O, as seen with \textit{Spitzer}, it is not completely an outlier, suggesting that other disks, such as those around MY Lup, IM Lup, DN Tau, and CX Tau are good candidates for the detection of $^{13}$CO$_{2}$. 
    
\end{enumerate}

Taken together, this study demonstrates that JWST-MIRI MRS has the ability to provide new and unique constraints on inner disk physical and chemical structures.

\begin{acknowledgements}
We thank the referee for thoughtful, constructive comments that improved the manuscript.
The following National and International Funding Agencies funded and supported the MIRI development: NASA; ESA; Belgian Science Policy Office (BELSPO); Centre Nationale d’Etudes Spatiales (CNES); Danish National Space Centre; Deutsches Zentrum fur Luft-und Raumfahrt (DLR); Enterprise Ireland; Ministerio De Econom\'ia y Competividad; Netherlands Research School for Astronomy (NOVA); Netherlands Organisation for Scientific Research (NWO); Science and Technology Facilities Council; Swiss Space Office; Swedish National Space Agency; and UK Space Agency.

E.v.D. acknowledges support from the ERC grant 101019751 MOLDISK and the Danish National Research Foundation through the Center of Excellence ``InterCat'' (DNRF150). B.T. is a Laureate of the Paris Region fellowship program (which is supported by the Ile-de-France Region) and has received funding under the  Marie Sklodowska-Curie grant agreement No. 945298. D.G. would like to thank the Research Foundation Flanders for co-financing the present research (grant number V435622N). T.H. and K.S. acknowledge support from the ERC Advanced Grant Origins 83 24 28. I.K., A.M.A., and E.v.D. acknowledge support from grant TOP-1614.001.751 from the Dutch Research Council (NWO). I.K. and J.K. acknowledge funding from H2020-MSCA-ITN-2019, grant no. 860470 (CHAMELEON). G.B. thanks the Deutsche Forschungsgemeinschaft (DFG) - grant 138 325594231, FOR 2634/2. O.A. and V.C. acknowledge funding from the Belgian F.R.S.-FNRS. I.A. and D.G. thank the European Space Agency (ESA) and the Belgian Federal Science Policy Office (BELSPO) for their support in the framework of the PRODEX Programme. D.B. has been funded by Spanish MCIN/AEI/10.13039/501100011033 grants PID2019-107061GB-C61 and No. MDM-2017-0737. A.C.G. has been supported by PRIN-INAF MAIN-STREAM 2017 and from PRIN-INAF 2019 (STRADE). T.P.R acknowledges support from ERC grant 743029 EASY. D.R.L. acknowledges support from Science Foundation Ireland (grant number 21/PATH-S/9339). L.C. acknowledges support by grant PIB2021-127718NB-I00, from the Spanish Ministry of Science and Innovation/State Agency of Research MCIN/AEI/10.13039/501100011033

\end{acknowledgements}

\begin{appendix}

\section{Comparison with \textit{Spitzer}-IRS}\label{sec: comparison with spitzer}
\begin{figure*}
    \centering
    \includegraphics[scale=0.55]{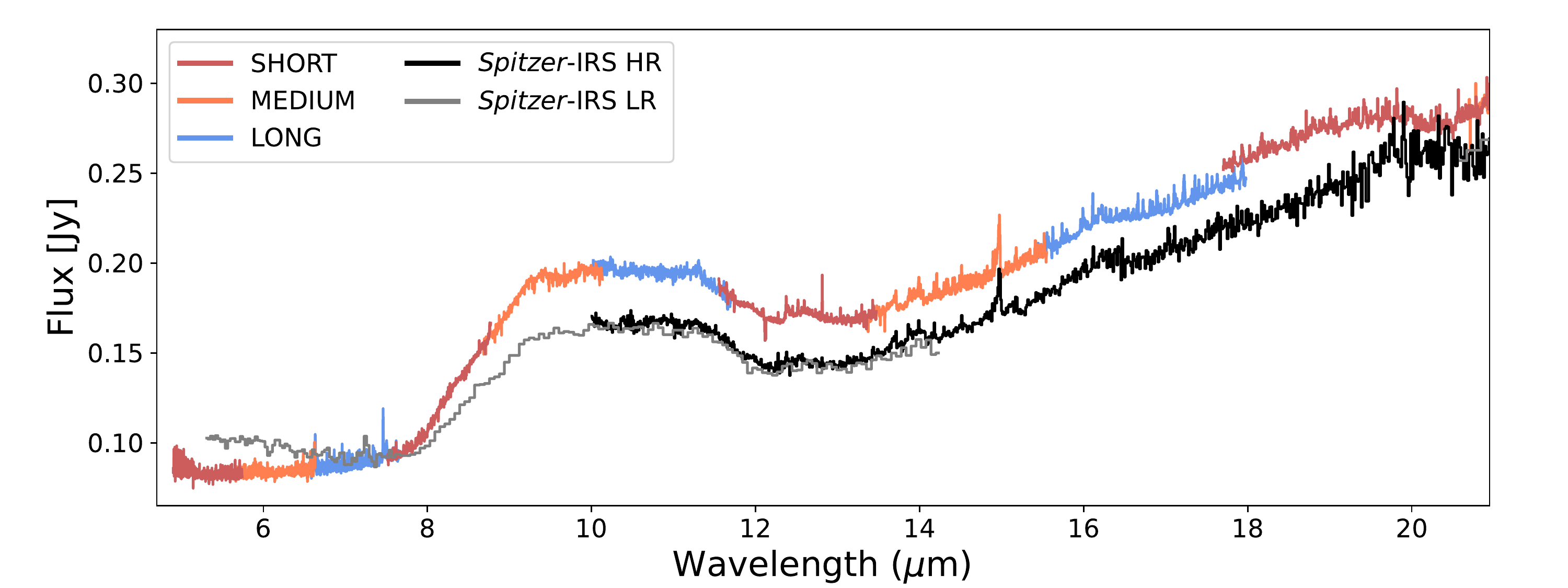}
    \caption{The JWST-MIRI MRS spectrum for GW Lup is shown with the sub-bands in different colors. The \textit{Spitzer}-IRS high-resolution (HR) and low-resolution (LR) data are shown for comparison. 
    }
    \label{fig: subbands and spitzer}
    
\end{figure*}

The comparison between the MIRI MRS data and the \textit{Spitzer}-IRS data for GW Lup is shown in Figure~\ref{fig: subbands and spitzer}. A spurious, single-pixel spike at 18.8 \mic\ has been removed from the MIRI spectrum, as in Figure~\ref{fig: spec}.

\color{black}
\section{Continuum subtraction}\label{sec: continuum subtraction}
\begin{figure*}
    \centering
    \includegraphics[scale=0.55]{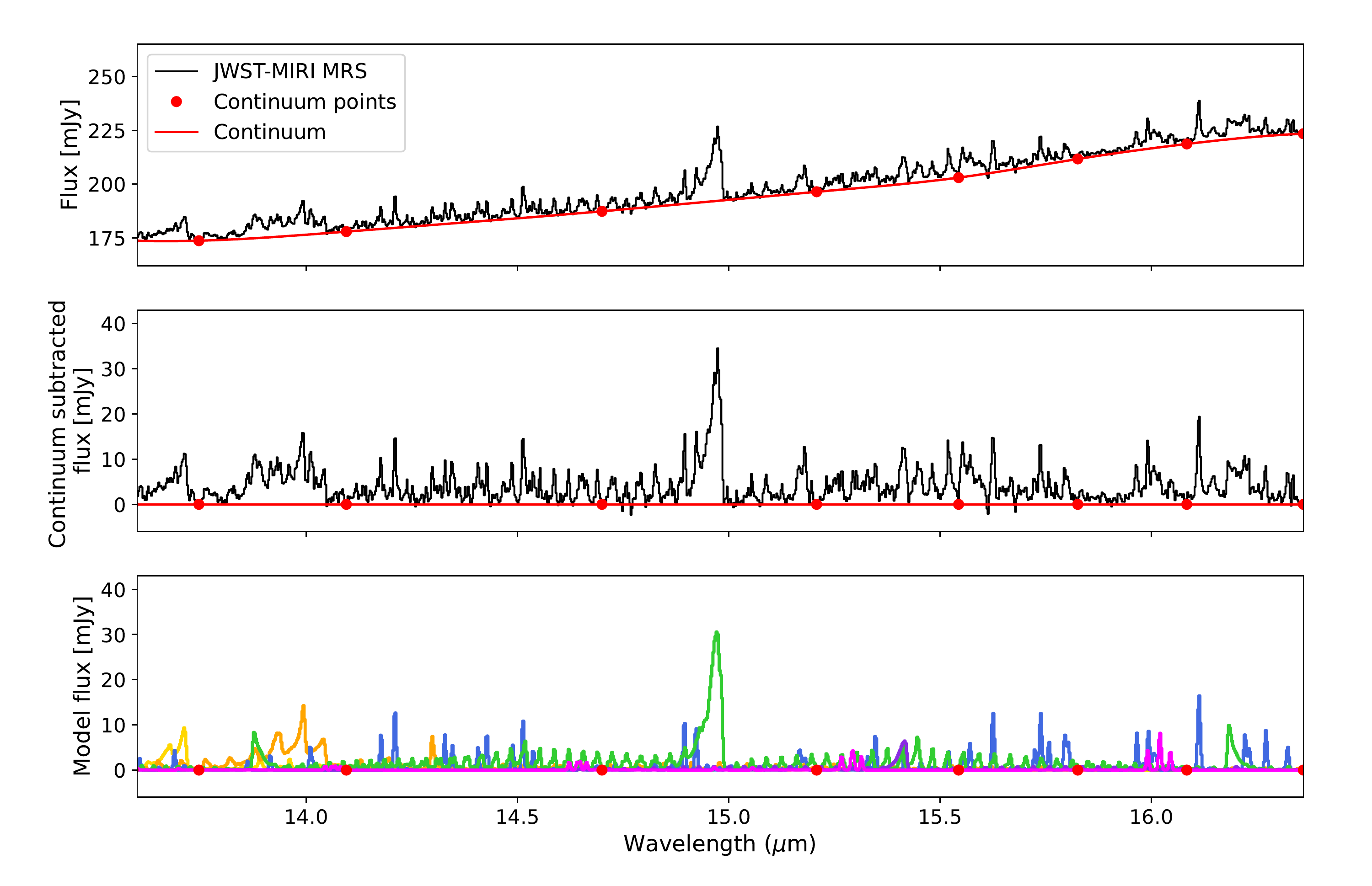}
    \caption{Top: The 13 to 16.3 \mic\ wavelength range of our JWST-MIRI MRS data of GW Lup (black). The continuum points that we select are shown as the red points and the interpolated continuum is shown as the red line. Middle: The continuum subtracted spectrum. Bottom: The best-fit models, as in Figure~\ref{fig: spec and model}, showing the continuum points relative to the molecular features. 
    }
    \label{fig: continuum subtraction}
    
\end{figure*}

The continuum is determined using a cubic spline interpolation (\texttt{scipy.interpolate.interp1d}) between selected regions with minimal line emission in the spectrum (Figure~\ref{fig: continuum subtraction}). Because the molecular emission is so rich in this wavelength region, the continuum points are selected to lie between emission features, which we confirm with our best-fit models (bottom panel). Due to the high signal-to-noise of the data and the fact that molecular features are not expected to produce an underlying continuum level at the column densities determined for this source, regions of low emission can be taken as the continuum level.

\section{$\chi^{2}$ procedure and maps}\label{sec: chi2 procedure and maps}
\begin{figure*}
    \centering
    \includegraphics[scale=0.75]{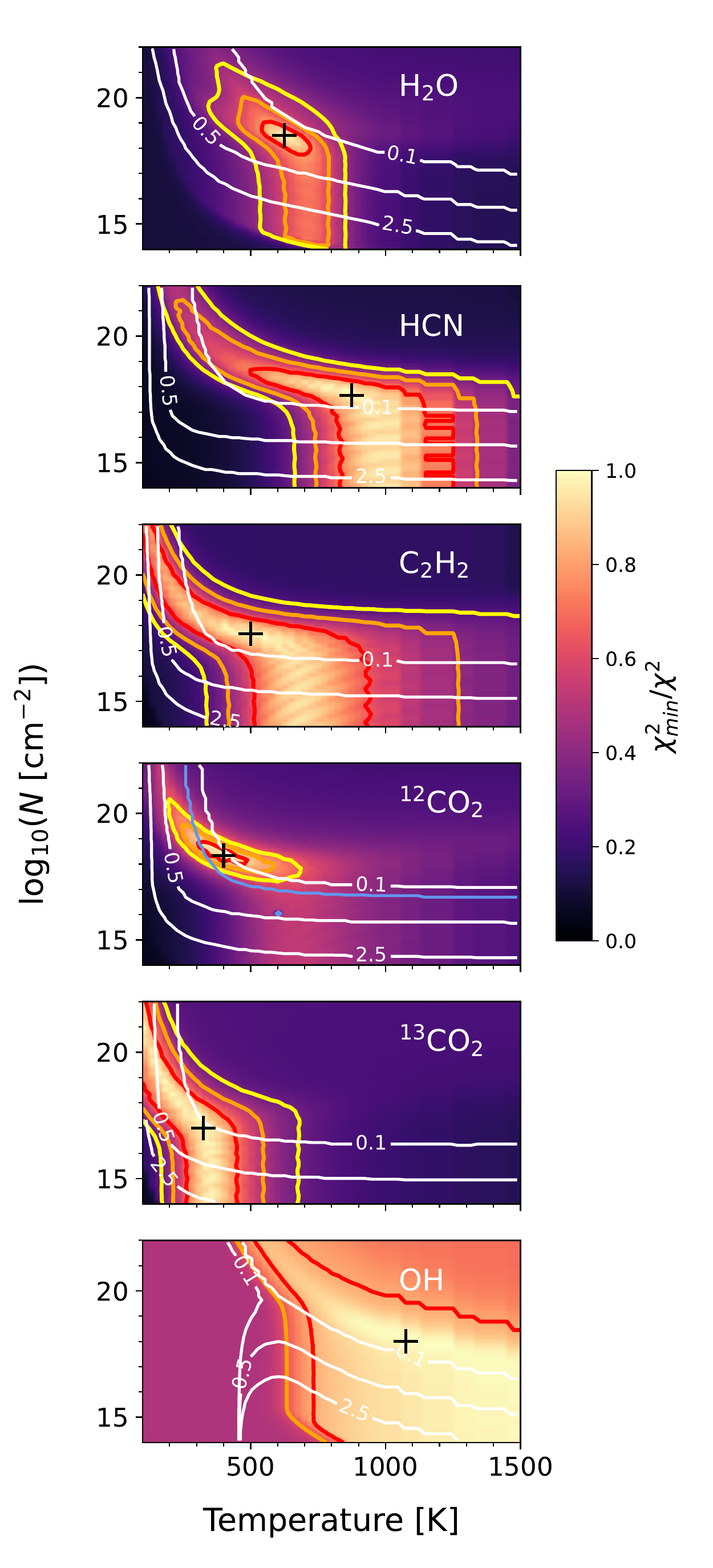}
    \caption{The $\chi^{2}$ maps for H$_{2}$O, HCN, C$_{2}$H$_{2}$, $^{12}$CO$_{2}$, $^{13}$CO$_{2}$, and OH (from top to bottom). The color-scale shows $\chi^{2}_{min}/\chi^2$. The red, orange, and yellow contours correspond to the 1$\sigma$, 2$\sigma$, and 3$\sigma$ levels. The white contours show the emitting radii in au, as given by the labels. The best-fit model is marked as the black plus. The best-fit model corresponds to $\chi_{min}^{2}$/$\chi^{2}$ = 1. These maps correspond to the $\chi^{2}$ and uncertainties after the third round in the iterative fitting procedure.
    The blue curve in the $^{12}$CO$_2$ plot is the best-fitting emitting radius for H$_2$O for comparison.}
    \label{fig: chi2 maps}
\end{figure*}

\begin{figure*}
    \centering
    \includegraphics[scale=0.45]{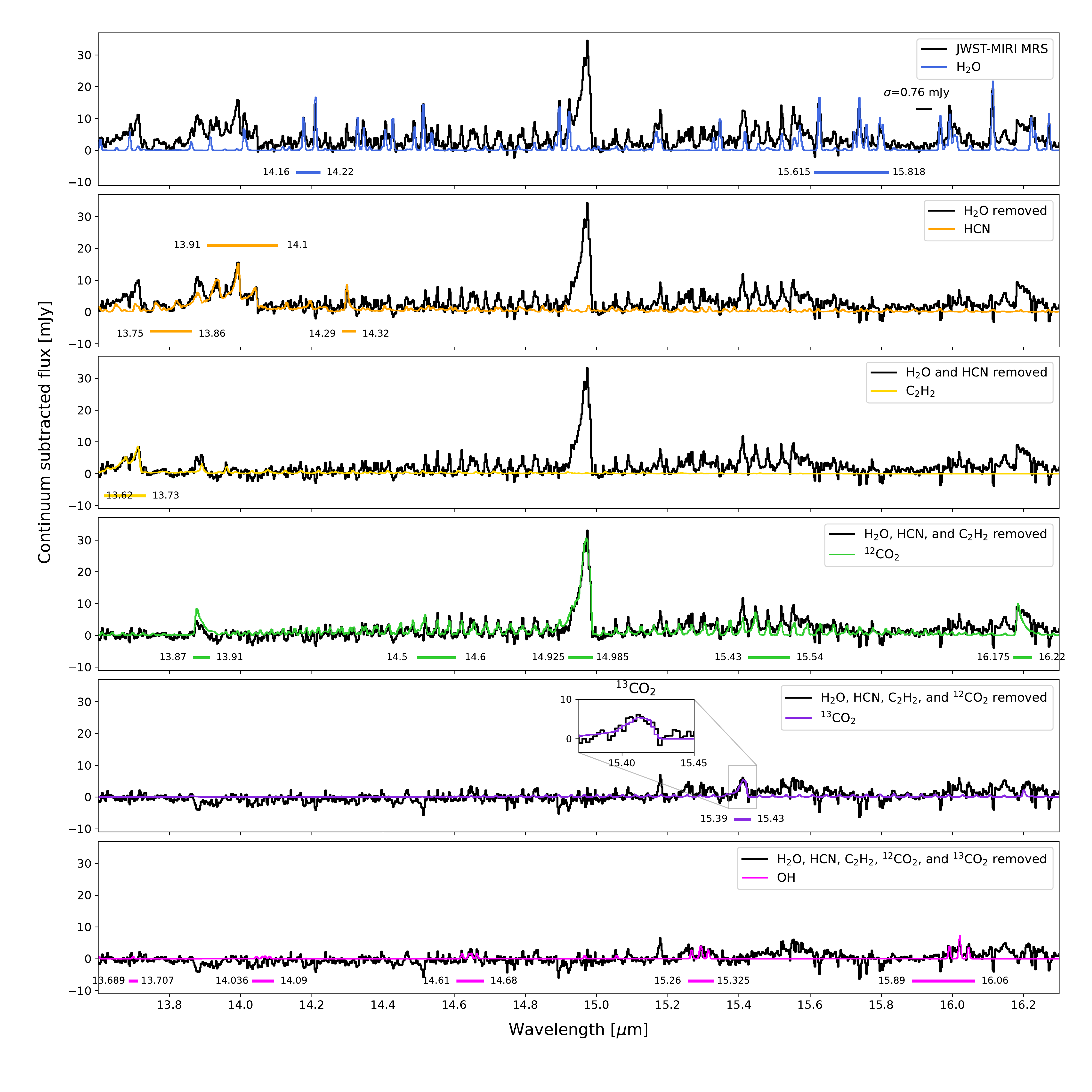}
    \caption{The best-fit model procedure is shown here. The top panel shows the best-fit H$_{2}$O model (blue) overlaid on the continuum-subtracted JWST-MIRI spectrum. In the second panel, the black spectrum is the observed spectrum after subtracting the H$_{2}$O model from the first panel. The best-fit HCN model is found using this as the input spectrum. This process continues down the panels. The spectral windows used for each species fit are shown as the horizontal bars, with the given starting and ending points. A region with minimal line emission from 15.90 to 15.94 \mic\ is chosen to determine the noise level (top panel). This region, before subtracting the models has a standard deviation of 0.76 mJy, however some low-level molecular emission is still present. 
    }
    \label{fig: fit procedure}
    
\end{figure*}

\begin{figure*}
    \centering
    \includegraphics[scale=0.45]{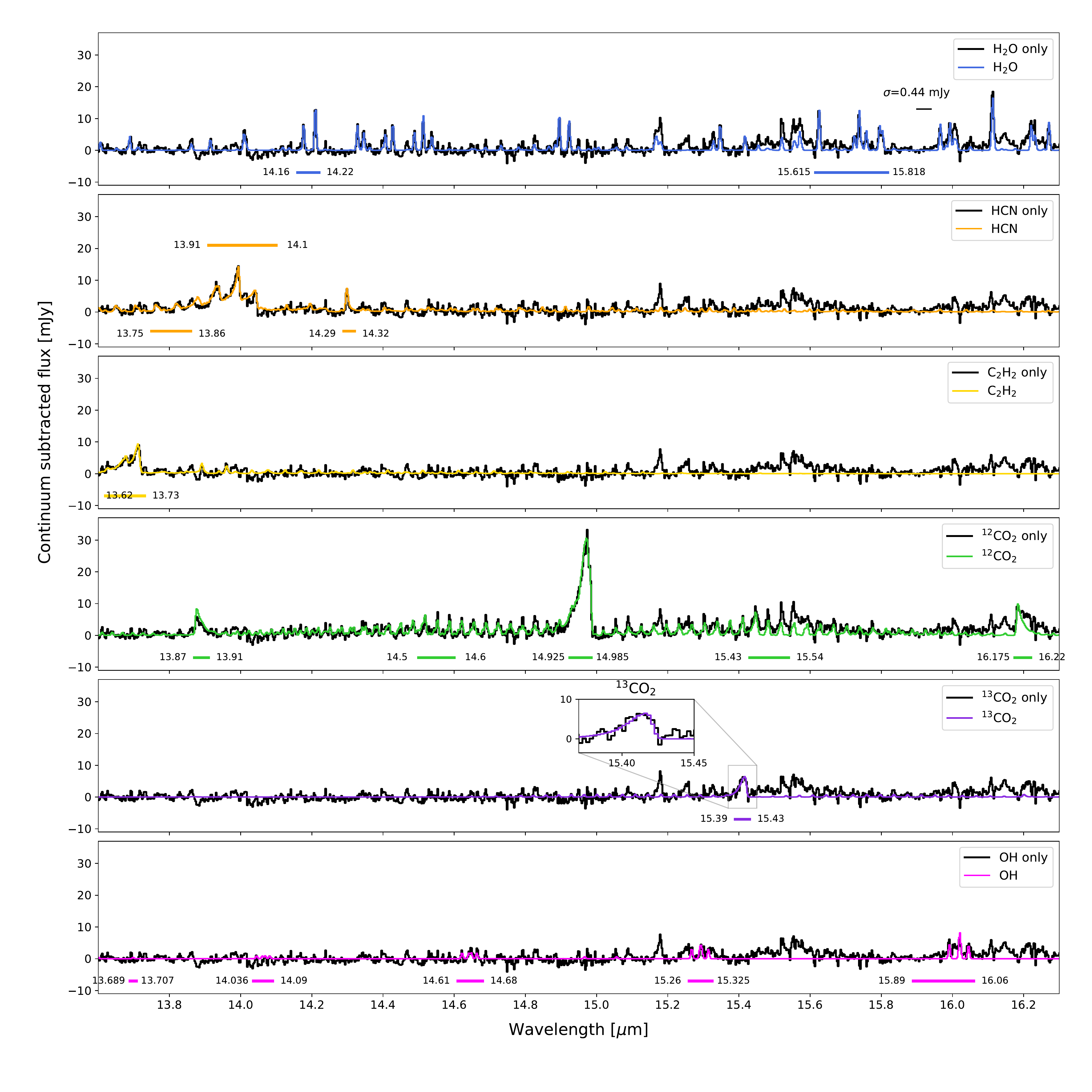}
    \caption{The same as Figure~\ref{fig: fit procedure}, but now showing the final best-fits after the iterative process. The spectra shown in black are the observed data after subtracting the best-fit models from the previous iterations for all molecules except that being fitted; these spectra are what is used in determining the best-fit for the species in each panel. The noise level is decreased from Figure~\ref{fig: fit procedure} because the excess emission, mostly from $^{12}$CO$_2$ in this region has been removed.
    }
    \label{fig: fit procedure 3} 
\end{figure*}

\begin{figure*}
    \centering
    \includegraphics[scale=0.5]{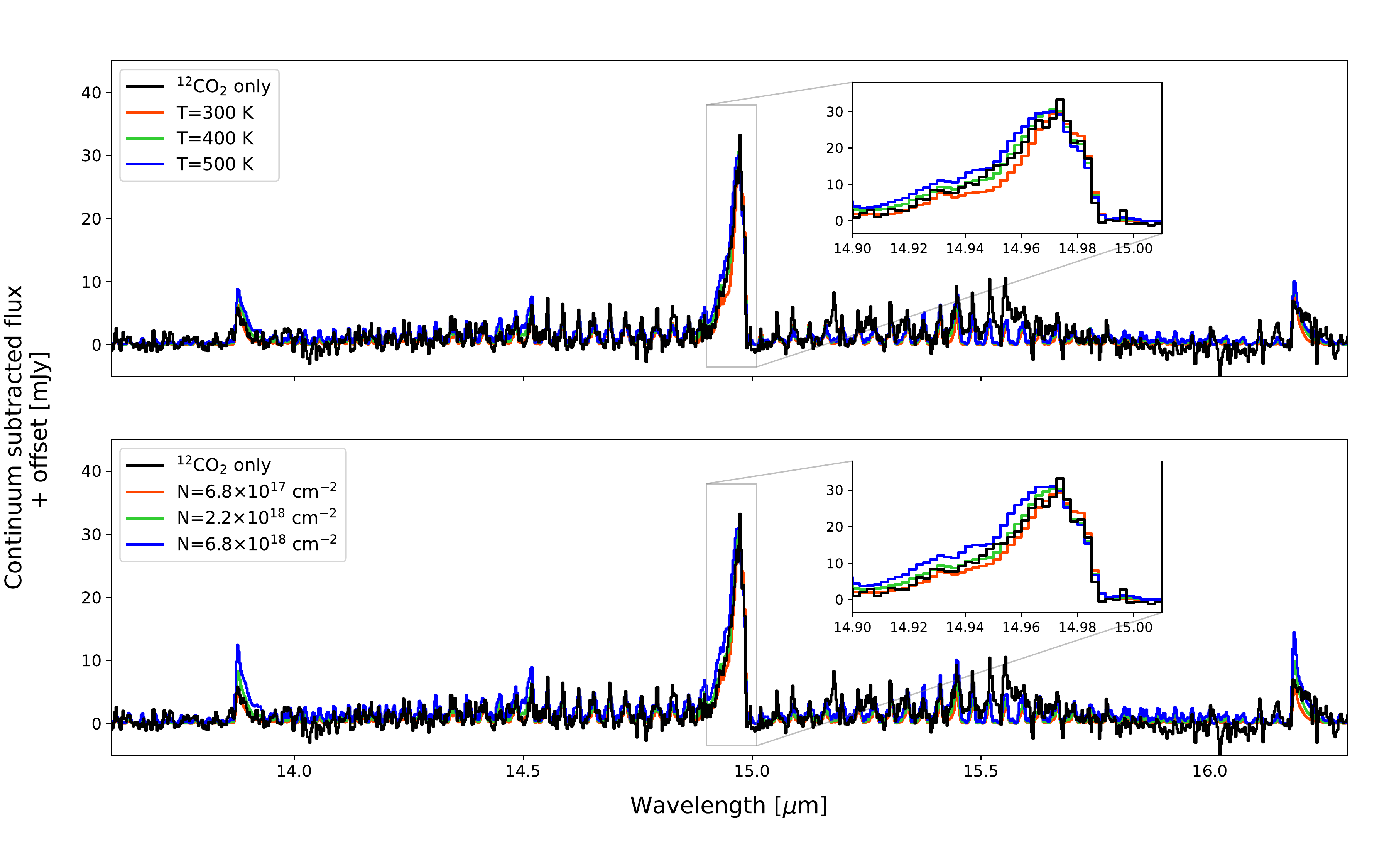}
    \caption{The $^{12}$CO$_2$-only data, as in Figure~\ref{fig: fit procedure 3}, with models showing the impact of temperature ($\pm$100 K; top) and column density ($\pm$0.5 dex; bottom). The emitting area for the models has been chosen to best match the data. The best-fit model found for $^{12}$CO$_2$, with a temperature of 400 K and a column density of 2.2$\times$10$^{18}$ cm$^{-2}$, is shown in green. The temperature controls the width of the $Q$-branches and the column density controls the position of the main $Q$-branch peak and the height of the hot-band peaks.}
    \label{fig: model example}
\end{figure*}

The reduced $\chi^{2}$ maps for H$_{2}$O, HCN, C$_{2}$H$_{2}$, $^{12}$CO$_{2}$, $^{13}$CO$_{2}$, and OH are shown in Figure~\ref{fig: chi2 maps}. The reduced $\chi^{2}$ is determined using the following formula: 
\begin{equation}
    \chi^2 = \frac{1}{N}\sum_{i=1}^{N} \frac{(F_{obs, i}-F_{mod, i})^2}{\sigma^2},
\end{equation}
where $N$ is the number of resolution elements in the spectral windows that the fit is done over and $\sigma$ is the standard deviation in a region with minimal line emission from 15.90 to 15.94 \mic\ (Figure~\ref{fig: fit procedure}). As the emitting radius is just a scaling factor, the degrees of freedom is only two for the column density and temperature. The contours in the reduced $\chi^2$ shown in Figure~\ref{fig: chi2 maps} are the 1$\sigma$, 2$\sigma$, and 3$\sigma$ levels determined as $\chi^2_{min}$ + 2.3, $\chi^2_{min}$ + 6.2, and $\chi^2_{min}$ + 11.8, respectively (see \citealt{numericalrecipesinc} and Table 1 and equation 6 of \citealt{avni76}). Any contribution from other species in this line-rich region of the spectrum increases the overall $\chi^2$ value, although the spectral windows are selected to minimize this contribution. The procedure is iterative, as described in Section~\ref{subsec: slab models}, to reduce the influence of overlapping molecular features on the best-fit parameters for a given species. The best-fit models are shown in Figure~\ref{fig: fit procedure 3}, along with the final noise level of 0.44 mJy. Figure~\ref{fig: model example} shows the effects of changing temperature and column density on the CO$_2$ model as an example.

\end{appendix}

\bibliographystyle{aasjournal}
\bibliography{biblio}

\end{document}